\documentclass{emulateapj}
\usepackage{apjfonts}

\usepackage{epsfig}

\def\deg{^{\circ}}
\def\P3hat{{\mathaccent 94 P}_3}

\def\A{{\bf A}~}
\def\B{{\bf B}~}
\def\An{{\bf A}}
\def\Bn{{\bf B}}
\def\pA{pulsar {\bf A}~}
\def\pB{pulsar {\bf B}~}
\def\pAn{pulsar {\bf A}}
\def\pBn{pulsar {\bf B}}

\shorttitle{Axis Orientations of J0737--3039 Pulsars}

\shortauthors{P. Demorest et al.}

\begin{document}

\title{Orientations of Spin and Magnetic Dipole Axes of Pulsars in the
J0737--3039 Binary Based on Polarimetry Observations 
at the Green Bank Telescope}

\author{P. Demorest$^1$, R. Ramachandran$^1$, D. C. Backer$^1$,
S. M. Ransom$^{2,3}$, V. Kaspi$^{2,3,4}$, J. Arons$^1$, A. Spitkovsky$^{1,5,6}$}

\affil{$^1$Department of Astronomy, University of California,
Berkeley, CA 94720-3411, USA \\ 
$^2$Department of Physics, McGill University, Montreal,
QC H3A 2T8, Canada}
\altaffiltext{3}{Center for Space Research, Massachusetts
Institute of Technology, Cambridge, MA 02139}
\altaffiltext{4}{Canada Research Chair, Steacie Fellow, CIAR
Fellow}
\altaffiltext{5} {KIPAC, Stanford University, P.O. Box 20450, MS 29, Stanford, CA 94309}
\altaffiltext{6} {Chandra Fellow}

\begin{abstract}
We report here the first polarimetric measurements of the pulsars in
the J0737--3039 binary neutron star system using the Green Bank
Telescope. We conclude both that the primary star (\pAn) has a wide hollow
cone of emission, which is an expected characteristic of the
relatively open magnetosphere given its short spin period, 
and that \A has a small angle between its
spin and magnetic dipole axes, $4\pm 3$ degrees. This near alignment
of axes suggests that \An's wind pressure on \pBn's
magnetosphere will depend on orbital phase.  This variable pressure is
one mechanism for the variation of flux and profile shape of \pB
with respect to the orbital phase that has been reported.  The
response of \pB to the \A wind pressure will also depend on the particular
side of its magnetosphere facing the wind at the spin phase when
\B is visible. This is a second  possible mechanism for variability.  
We suggest that \pB may have
its spin axis aligned with the orbital angular momentum owing to
\An's wind torque that contributes to its spindown.  Monitoring
the pulsars while geodetic precession changes spin orientations 
will provide essential evidence to test detailed theoretical models.
We determine
the Rotation Measures of the two stars to be $-112.3\pm 1.5$ and
$-118\pm 12$ rad m$^{-2}$. \end{abstract}


\section{Introduction}
\label{sec-intro}
The double pulsar system, PSR J0737--3039, recently reported by Burgay
et al. (2003) and Lyne et al. (2004) is likely to unlock many mysteries
concerning isolated neutron star magnetospheres and winds as well as the
nature of the pulsar emission mechanism. A 22.7-ms pulsar (\An) and a
2.77 sec pulsar (\Bn) revolve about each other in a 2.4h, nearly
edge-on orbit.  
The two stars show interesting and contrasting emission properties.
Pulsar \A exhibits a complex double profile structure. Apart from an
eclipse when \A is behind \B (with respect to our sight line),
this pulsar does not show significant variation in its flux. 
The flux and profile structure of \B show remarkable variations
as a function of orbital phase (Lyne et al. 2004).


A simple pressure balance calculation leads to the conclusion 
that the MHD wind of \pA will surpress a large fraction 
of the quasi-static \B magnetosphere.  The eclipse duration of \pA 
is roughly consistent with this idea (Lyne et al. 2004; 
Kaspi et al. 2004). Detailed calculations of the \An-wind/\Bn-magnetosphere
interactions are required for comparisons with observations
(Arons et al. 2004). In this model a bow shock, magnetosheath, 
magnetopause and magnetotail structures, 
which are relativistic analogs of solar wind/Earth magnetosphere structures,
are established around \pBn.

In this work, we first present polarimetric measurements on these
two stars. These observations were conducted at the National Radio
Astronomy Observatory\footnote{The National Radio Astronomy Observatory
(NRAO) is owned and operated by Associated Universities, Inc under
contract with the National Science Foundation.} Green Bank Telescope
(GBT). After describing our observational setup in
\S\ref{sec-observation}, we present our results in \S\ref{sec-result}.
Our observations place strong constraints on the
geometrical orientation of {\bf A}'s rotation and magnetic axes. We also
present our Rotation Measure (RM) measurements for the two stars. In 
\S\ref{sec-model} we discuss implications of our
polarimetric results within the context of the Arons et al. model.

\section{Observations}
\label{sec-observation}
Observations were performed at the GBT with two backends: BCPM
(Berkeley--Caltech Pulsar Machine; see e.g., Camilo et al. 2002) and 
GBPP (Green Bank--Berkeley Pulsar Processor; see e.g., Backer, Wong \&
Valanju 2000). The pulsar system was
observed at four different frequencies on 2003 December 11, 19, 23,
\& 24, and 2004 January 1. The bandwidths of our BCPM observations were
96 MHz at 2.2 GHz and 1.4 GHz, and 48 MHz at 0.82 GHz and  0.43 GHz.
The GBPP bandwidths were 28 MHz at 2.2 GHz, 1.4 GHz
and 0.82 GHz and 11 MHz at 0.43 GHz. 

On 2003 December 11 we observed \A with the GBPP at 1.4 GHz using
the published ephemeris (Burgay et al. 2003) and an integration time of
100s.  The GBPP provides synchronously averaged, coherently dedispersed
profiles that allow calculation of all Stokes parameters.

We first detected and then generated an ephemeris for \B with the
BCPM filter bank total-power data of 2003 December 11.  This ephemeris
was used for GBPP observations of \B at 0.82 GHz on 2003 December 19.
GBPP observations continued with \A at 0.43 GHz (briefly), 0.82 GHz and 2.2 GHz
and \B at 0.43 GHz. We also measured
a correlated (fully linearly polarized)  noise
source and observed pulsars with known
polarization characteristics (PSR B0656+14 and PSR B0919+06) for
calibration purposes. GBPP observations of PSR B1929+10 at 0.82 GHz by I.
Stairs for another program on 2003 December 26 were also analyzed for
calibration.

\section{Results}
\label{sec-result}
\begin{figure}[t]
\begin{center}
\epsfig{file=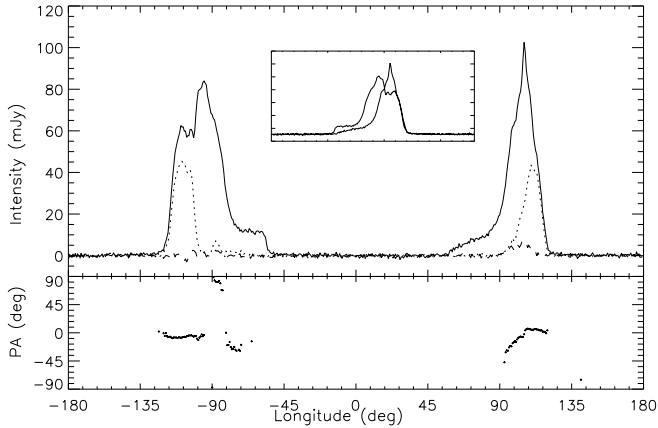,height=6cm}
\caption{
Polarization characteristics of J0737--3039A at 0.82 GHz.  The inset
gives the two portions of the profile `mirror-folded' to show similarity
of outer edges. See text for details.
}
\label{fig:avgA}
\end{center}
\end{figure}

The top panel of Fig.~\ref{fig:avgA} shows the average profile of \A
at 0.82 GHz. The pulsar shows two prominent regions of emission with
sharp outer edges, and shallower inner edges. The dotted and dashed
lines indicate observed average profiles in linear and circular
polarization, respectively. We also show the linear polarization
position angle as a function of pulse longitude in the bottom panel of
Fig.~\ref{fig:avgA}. The longitude region $-100\deg$ to $-70\deg$, and
probably the region from $+90\deg$ to $+105\deg$, are affected
by the presence of quasi-orthogonal polarization modes.

The two emission regions display rough mirror symmetry in both total
intensity and linear polarization. We show
in the inset on the figure two portions of the profile ``mirror-folded''
to align the outer edges and emphasize this point.
These mirror symmetries suggest to us that
these two portions of the pulse arise from traversals of a wide-angle
hollow cone at the same magnetic pole
rather than the two opposite poles. 
We have chosen the arrangement of the profile
in rotational longitude with steep outer edges in Fig.~\ref{fig:avgA}
owing to what is often seen in more compact hollow cones (e.g.,
Manchester 1996). 

\begin{figure}[t]
\begin{center}
\epsfig{file=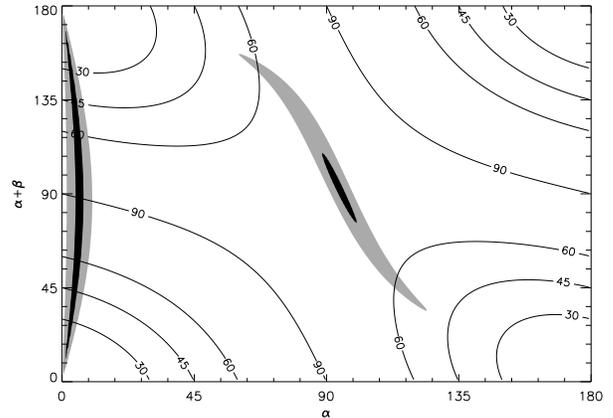,height=6cm}
\caption{
$\chi^2$ contours for the RVM fit for \A.  The filled contours represent
1$\sigma$ and 3$\sigma$ confidence limits.  The line contours show
the emission cone opening half-angle $\rho$ for each solution.
}
\label{fig:rvmfit}
\end{center}
\end{figure}

\begin{figure}[t]
\begin{center}
\epsfig{file=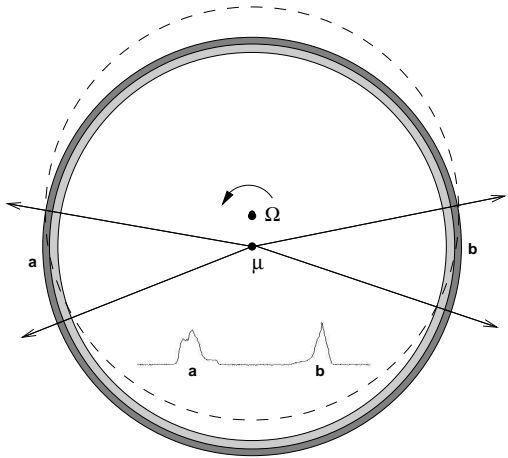,height=6cm}
\caption{The proposed geometry of the emission beam of PSR J0737--3039A,
  as viewed from above the pole. 
  The grayscale circles indicate the emission cone,
  centered around the magnetic axis ($\vec{\mu}$). The rotation axis
  is given by $\vec{\Omega}$. 
  The `dash'
  line indicates the line of sight trajectory around $\vec{\Omega}$.
  The radiating vectors from $\vec{\mu}$ are projections of field
  lines diverging from the pole which nominally give the PA
  of linear polarization. The two emission regions in the profile
  are marked as {\bf a} and {\bf b}.}
\label{fig:schematic}
\end{center}
\end{figure}

Assuming the rotating vector model (RVM; Radhakrishnan \& Cooke 1969), we
performed a least squares fit to the position angle sweep to obtain the
value of the angle ($\alpha$) between $\vec{\mu}$ and $\vec{\Omega}$  and
the angle ($\beta$) between the line of sight and $\vec{\mu}$  at closest approach.
For this purpose, we ignored the longitudes mentioned above
where the PA is likely corrupted by quasi-orthogonal emission modes. 
The fit was done via a simple grid search in $\alpha$
and $\beta$ (e.g. Nice et al. 2001).  This allowed us to map out
$\chi^2$ contours for the solutions, which are plotted in
Fig.~\ref{fig:rvmfit}.  Two solutions appear, one with $\alpha \sim
5\deg$ and one with $\alpha \sim 90\deg$.  The large-$\alpha$ solution
is unrealistic since it requires a beam opening angle $2\rho \sim
180\deg$, a geometry where the RVM is likely to be invalid.
Furthermore, in this case we would expect to see emission from both
magnetic poles, which is inconsistent with the mirror symmetric profile.
We favor the small-$\alpha$ solution since it avoids both these problems
(see Fig.~\ref{fig:schematic} for a diagram of this geometry).
Unfortunately, this leaves $\beta$ completely unconstrained.  Measuring
$\beta$ would give a lower limit on the misalignment between
$\vec{\Omega}$ and the orbital angular momentum, since the orbital
inclination is well known.  However, in the small-$\alpha$ solution
$\beta \sim \rho$, so we might expect $\beta \sim 50\deg$ based on
opening angles of other millisecond pulsars. 


\begin{figure*}[t]
\begin{center}
\epsfig{file=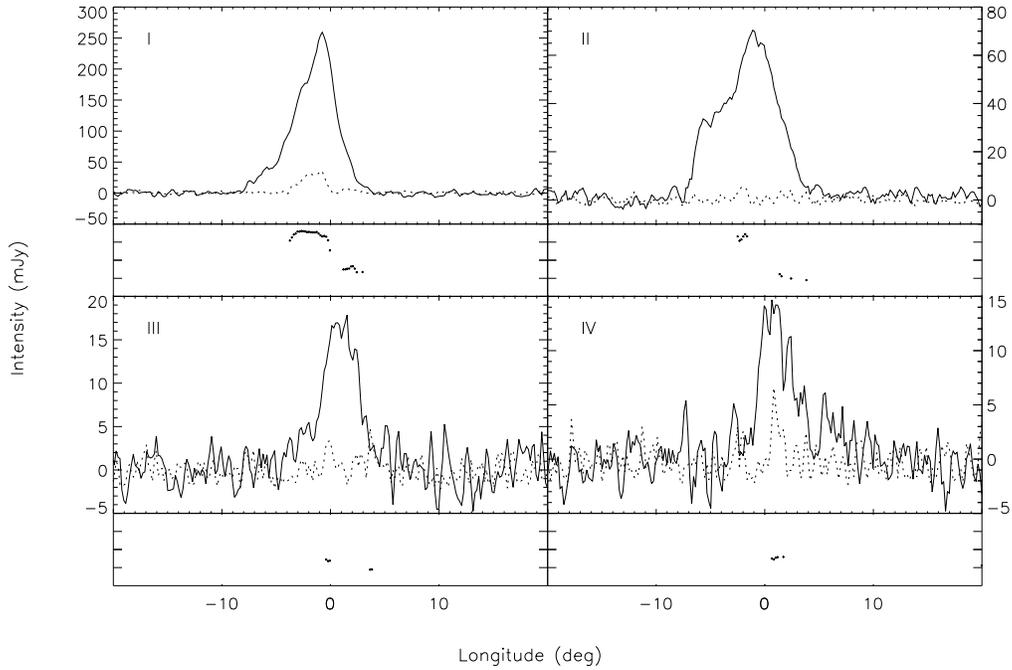,height=10cm}
\caption{Polarization characteristic of J0737--3039B. The four panels
give the results for the four orbital phase ``windows'' see text for
details.}
\label{fig:avgB}
\end{center}
\end{figure*}

The characteristics of {\bf B}'s emission are much more complex.  Lyne et
al. (2004) show that the amplitude and pulse profile vary with orbital
phase. Our sensitive data exhibit local maxima in four different
``windows'' in the orbit ($I$ to $IV$\footnote{with the convention
followed by Lyne et al. (2004), the ranges of orbital phases (true
anomaly) we have taken for the four windows are: $186\deg - 216\deg$,
$252\deg - 300\deg$, $336\deg - 30\deg$ and $90\deg - 132\deg$,
respectively}). Within the limit of our sensitivity, this 4-window
behavior and details of profile evolution within the windows are
frequency independent over 0.43 GHz to 1.4 GHz. 
The signal is seen in almost all other
orbital phases at very low flux levels except for a phase range of $30\deg$
to $92\deg$, where no flux is detected above 0.1\% of that in window $I$. 
The centroid of the strongest emission (windows $I$ \& $II$)
is $\sim 30\deg$ prior to when \A is behind \Bn, and the centroid
of the phase range when \B is invisible is $\sim 30\deg$ prior to
when \A is in front of \B. 
Due to the pulse structure evolution, we
present the results separately for the four windows in
Fig.~\ref{fig:avgB} from our 0.82-GHz observation with the same
conventions as in Fig.~\ref{fig:avgA}. In window 1 the pulse
is polarized at the 15\% level and shows 
orthogonal polarization modes, while in window 2 the polarization
is less than a few per cent.
A detailed study of individual pulses from \B and their evolution as
a function of orbital phase by Ramachandran et al. (2004) concludes that
the subpulse fluctuations are similar to those found 
in other pulsars of similar period and pulse morphology. 

\subsection{Rotation Measures}
\label{sec-RM}
We have measured the RM of the two pulsars from our 0.82 GHz
observations.  Assuming linearity of polarization position angle sweep
within the small bandwidth of 28 MHz, 
we have derived RM values for \pAn and \pB from window $I$. We have
obtained consistent values of $-112.3 \pm 1.5$ rad m$^{-2}$ and $-118\pm
12$ rad m$^{-2}$ towards \A \& \Bn, respectively. These values are
not corrected for ionospheric contributions. 
Within our measurement errors, we do not see significant
orbital phase dependent variations. 

\section{Discussion}
\label{sec-model}


A remarkable feature of both the Lyne et al. (2004) and our GBT
observations of  J0737--3039B is the dependence of  flux and pulse
profile, including polarization properties, on orbital phase, but not on
radio frequency. We concur in general with Lyne et al. that the variable
influence of \pA on the magnetosphere of \B is the likely 
source of the changes. In a companion paper (Arons et al. 2004) we
develop a detailed model in which \An's MHD wind confines the
\B magnetosphere. We outline this model below, and make
connections between observational features that both stimulated the
model making and provide directions for future work.

In our model
the MHD wind from \A decelerates through a relativistic bow shock that
envelops \Bn's magnetosphere. Between this shock and the boundary of
\Bn's closed magnetosphere (\Bn's magnetopause) the shocked \A wind plasma
forms a relativistic hot layer whose optical depth to synchrotron
absorption at 500 MHz is at least 100, and may be as high as 5000. 
This layer (the magnetosheath) is the likely cause of the eclipse
of \A by \Bn. 
In this model we further interpret the faintest
region of \B emission near its superior conjunction as the result
of absorption by the same magnetosheath layer that in these
phases is between the observer and the pulsar emission region 
deep within the \B magnetosphere. 
In Kaspi et al. (2004) we report that the \A eclipse is asymmetric 
with slower flux decrease on ingress and a more
rapid recovery on egress. As discussed above the strongest and faintest regions
of \B emission are also asymmetric: both precede times
of conjunction by $\sim 30^\circ$. In our model these are attributed
to the prograde rotation of \B that leads to an asymmetric magnetopause
and magnetotail.
The frequency independence of the eclipse light curve 
over our observing bands is attributed to sharp boundaries, high 
optical depths and partial covering in space and/or time. 

We favor the interpretation of our polarization observations  of \pA
that has the dipole field axis nearly aligned with the spin
axis, which itself is oblique to the orbital plane and the line of
sight. 
Pulsar \B would then experience a significant difference in wind pressure 
and content around the orbit.  This variable pressure will affect the size
of the polar cap and current structure
that could lead to the \B variations.


The \A wind, asymmetric or not, will produce a propeller torque on
\B owing to \Bn's rotation.  This torque contributes to the
observed $\dot{P}$ of \Bn.  In this model the \B emission results from
voltage and $e\pm$-avalanche current deep within its magnetosphere
similar to that in normal pulsars. The ``normalcy'' of \B emission is
supported by these polarization observations as well as the single
pulse study described elsewhere (Ramachandran et al.  2004).

When we observe \pBn, it presents a different face of its closed and
confined magnetosphere to the \A wind depending on the orbital phase.
This variable orientation will also change the \B magnetosphere and,
we suggest, its polar cap and the beamed emission we detect. At this
point we cannot distinguish between these two possible sources of
modification of the secondary star -- variable \A wind pressure
impinging on \Bn, and variable \B internal structure at the rotation
phase of observation. They both are seemingly consistent with the
frequency independence of the reported phenomena. Both cases will lead
to variations in the polar cap size and relativistic current structure
of \Bn, which will affect the observed flux and its pulse morphology.


In the case of the variable wind pressure, the phase of modulation of
\B is expected to precess around the orbit as the spin of \A undergoes
geodetic precession. Geodetic precession of \B will have a more
complex effect on its internal response to a wind pressure at the spin
phase of observation owing to the multiplicity of angles involved.


Our proposed geometry of \An's beam will be testable as
geodetic precession moves the observer through the cone and into
a region of invisibility during the 70-year cycle (Lyne et
al. 2004). 
Is it improbable that we are seeing both pulsars
at the same time? We suggest that the \A wind torque on \B has aligned
pulsar \Bn's spin axis with the orbital angular momentum over
time. In this case \B would need to have its dipole axis at
$\alpha\sim 90^\circ$, and therefore we will continue to view \B
independent of geodetic precession. An interesting question is
whether the torque of the \A wind can misalign the dipole while
aligning the spin. The full history of the evolution of the system and
the dueling magnetospheric winds remains to be written. In its infancy
\B might have had a short spin period and strong magnetic field. If
so, then the early wind from \B could have dominated the \A
magnetosphere and altered its spin and magnetic dipole.

\subsection{Summary}
\label{sec:sum}
Our polarimetric observations indidate that pulsar {\bf A}'s spin and
magnetic axes are nearly aligned. This leads us to consider two
possible mechanisms for the variability of \B with orbital phase --
(1) a pole to equator asymmetry of the \A wind, and (2) a rotational
asymmetry of the force balance radius of the \B magnetosphere as it
adjusts to the \A wind.
The interaction also contributes to the spin-down torque on the \B pulsar
which may also lead to alignment of the \B spin and orbital momentum
vectors. Geodetic precession will provide an important means of
separating the relative importance of the various effects in the
coming years. Theoretical calculations are underway to explore these
ideas quantitatively.

\acknowledgements 

We thank the GBT staff, and in particular Carl
Bignell, Frank Ghigo, Glen Langston and Karen O'Neil, for extensive
help with the observations and very useful discussions. We thank the
initial investigators of the J0737-3039 system for propagating news of
this wonderful discovery in advance of publication. 
We thank the organizers of the 2004 January Aspen Center
for Physics workshop for providing time to present these results that
were developed prior to and indepently of the meeting.


\begin{thebibliography}{}
\bibitem[]{} Arons, J.A., Spitkovsky, A., et al. 2004, ApJ, in preparation
\bibitem[]{} Backer, D.C., Wong, T. \& Valanju, J. 2000, ApJ, 543, 740
\bibitem[]{} Burgay, M., D'Amico, N., Possenti, A. et al. 2003,
             Nature, 426, 531
\bibitem[]{} Camilo, F., Stairs, I. H., Lorimer, D. R. et al. 2002,
ApJ, 571, 41
\bibitem[]{} Kaspi, V., Ransom, S., Backer, D. C. et al. 2004, 
    ApJ, Submitted (astro-ph/0401614)
\bibitem[]{} Lyne, A. G., Burgay, M., Kramer, M. et al. 2004, Science, in press 
   (astro-ph/0401086)
\bibitem[]{} Nice, D.J., Splaver, E.M. \& Stairs, I.H. 2001, ApJ, 549,
   516
\bibitem[]{} Radhakrishnan, V., Cooke, D. 1969, ApL, 3, 225
\bibitem[]{} Ramachandran, R., Backer, D. C., Demorest, P. et al. 
    2004, in preparation
\end{thebibliography}
\end{document}